\documentclass{IEEEcsmag}

\usepackage[colorlinks,urlcolor=blue,linkcolor=blue,citecolor=blue]{hyperref}
\expandafter\def\expandafter\UrlBreaks\expandafter{\UrlBreaks\do\/\do\*\do\-\do\~\do\'\do\"\do\-}
\usepackage{upmath,color}

\usepackage{background}
\usepackage{tikzpagenodes}
\backgroundsetup%
{   angle=0,
	scale=1,
	contents=%
	{   \begin{tikzpicture}[remember picture,overlay, scale=1]
			\node[anchor=north,text=gray,text opacity=1,xshift=2.25cm,yshift=2.75cm,font=\scriptsize ,align=center] at (current page.north) {This article has been accepted for publication in IEEE Computer Graphics and Applications. This is the author's version which has not been fully edited and \\ content may change prior to final publication. Citation information: DOI 10.1109/MCG.2024.3447775};        
			\node[anchor=south,text=gray,text opacity=1,xshift=2.25cm,yshift=2.9cm,font=\scriptsize ,align=center] at (current page.south) {\textcopyright~2024 IEEE. Personal use is permitted, but republication/redistribution requires IEEE permission. See https://www.ieee.org/publications/rights/index.html for more information.};  
		\end{tikzpicture}
	}
}

\jvol{}
\jnum{}
\paper{}
\jmonth{August}
\jname{}
\jtitle{}
\pubyear{2024}

\usepackage{glossaries}
\glsdisablehyper
\newacronym{NISQ}{NISQ}{noisy intermediate-scale quantum}
\newacronym{WFC}{WFC}{ave function collapse} 
\newacronym{PCG}{PCG}{procedural content generation}
\newacronym{PIPCG}{PIPCG}{probabilistic iterative PCG}
\newacronym{CWFC}{CWFC}{classical wave function collapse}
\newacronym{QWFC}{QWFC}{quantum wave function collapse}
\newacronym{HWFC}{HWFC}{hybrid quantum-classical wave function collapse}
\newacronym{QRNG}{QRNG}{quantum random number generator}
\newacronym{QPCG}{QPCG}{quantum procedural content generation}

\usepackage{xspace}
\newcommand{\ie}{i.e.\@\xspace}
\newcommand{\eg}{e.g.\@\xspace}
\newcommand{\cf}{cf.\@\xspace}

\newcommand{\kolkata}{Falcon r5.11 processor version 1.14.8, 27 qubits}
\newcommand{\kyoto}{Eagle r3 processor version 1.2.4, 127 qubits}
\newcommand{\osaka}{Eagle r3 processor version 1.0.3, 127 qubits}

\usepackage{braket}
\usepackage{pgfplotstable}
\usepackage{tikz}
\usetikzlibrary{calc}
\usetikzlibrary{decorations.pathreplacing}
\usetikzlibrary{patterns}
\usetikzlibrary{shapes.geometric}
\pgfplotsset{compat=newest}

\renewcommand{\sfdefault}{phv}
\renewcommand{\rmdefault}{phv}

\usepackage{sfmath}
\newcommand{\standardfont}{\sffamily{\fontsize{10pt}{12}\selectfont}}
\tikzset{font=\standardfont}

\usepackage{amsmath,amssymb,bbm}
\DeclareMathOperator*{\argmin}{arg\,min}
\newcommand{\rand}[1]{\boldsymbol{#1}}
\newcommand{\operator}[1]{#1}

\definecolor{colorA}{HTML}{006BA3} 
\definecolor{colorB}{HTML}{FAB24A} 
\definecolor{colorC}{HTML}{619B8A} 
\definecolor{colorD}{HTML}{A56FEB} 
\definecolor{colorE}{HTML}{F4989C} 
\definecolor{colorF}{HTML}{93032e} 
\colorlet{colorAl}{colorA!25!white}
\colorlet{colorBl}{colorB!25!white}
\colorlet{colorCl}{colorC!25!white}
\colorlet{colorDl}{colorD!25!white}
\colorlet{colorEl}{colorE!25!white}
\colorlet{colorFl}{colorF!25!white}

\tikzstyle{subfiglabel}=[font=\standardfont\bfseries]
\tikzstyle{infonode}=[draw,ultra thick,rounded corners=.3cm]
\tikzstyle{infoline}=[draw,thick,->]
\tikzstyle{index}=[draw,thick,rounded corners=.1cm,colorA,fill=white,text=black,inner sep=0]
\tikzstyle{value}=[draw,thick,rounded corners=.1cm,colorB,fill=white,text=black,inner sep=0]
\tikzstyle{segment}=[draw,thick,dotted,rounded corners=.1cm,colorC]
\tikzstyle{basictile}=[index,minimum width=.65cm,minimum height=.65cm]
\tikzstyle{tileA}=[pattern=crosshatch, pattern color=colorA]
\tikzstyle{tileB}=[pattern=north west lines, pattern color=colorAl]
\tikzstyle{tiletext}=[font=\standardfont\scriptsize,inner sep=0pt]
\tikzstyle{signaltile}=[index,minimum width=.65cm,minimum height=.65cm,ultra thick,densely dotted,draw=colorF,fill=none]
\tikzstyle{circuitline}=[draw,colorA]
\tikzstyle{circuitbigline}=[draw,very thick,colorA]
\tikzstyle{pngatec}=[draw,semithick,circle,fill=black,draw=black,minimum width=.2cm,inner sep=0pt]
\tikzstyle{pngatet}=[draw,semithick,fill=colorBl,regular polygon,regular polygon sides=6,rounded corners=.05cm,minimum width=.75cm,minimum height=.55cm,inner sep=0pt,font=\standardfont\scriptsize]
\tikzstyle{pngatel}=[semithick]
\tikzstyle{pngatep}=[draw,semithick,fill=colorBl,rounded corners=.05cm,minimum width=.45cm,minimum height=.45cm,inner sep=0pt,font=\standardfont\scriptsize]

\let\labelindent\relax
\usepackage{siunitx}
\usepackage{enumitem}
\usepackage[mode=buildnew]{standalone}
\usepackage{cleveref}

\newlist{Scondenum}{enumerate}{1}
\setlist[Scondenum,1]{label=S\arabic*), ref=S\arabic*}
\creflabelformat{Scondenumi}{#2#1#3}
\crefname{Scondenumi}{condition}{conditions}
\newlist{Vcondenum}{enumerate}{1}
\setlist[Vcondenum,1]{label=V\arabic*), ref=V\arabic*}
\creflabelformat{Vcondenumi}{#2#1#3}
\crefname{Vcondenumi}{condition}{conditions}

\crefname{figure}{Figure}{Figures}
\crefname{equation}{Equation}{Equations}

\begin{document}
	
	\sptitle{FEATURE ARTICLE}
	
	\title{Quantum Wave Function Collapse for Procedural Content Generation}
	
	\author{Raoul Heese}
	\affil{Fraunhofer ITWM, Fraunhofer-Platz 1, 67663 Kaiserslautern, Germany}
	
	\markboth{FEATURE}{FEATURE}
	
	\begin{abstract}\looseness-1
		Quantum computers exhibit an inherent randomness, so it seems natural to consider them for procedural content generation. In this work, a quantum version of the famous (classical) wave function collapse algorithm is proposed. This quantum wave function collapse algorithm is based on the idea that a quantum circuit can be prepared in such a way that it acts as a special-purpose random generator for content of a desired form. The proposed method is presented theoretically and investigated experimentally on simulators and IBM Quantum devices.
	\end{abstract}
	
	\maketitle
	
	\chapteri{W}\gls{WFC} is a powerful tool for \gls{PCG} that is for example used in video games, as it can save significant development time by automating the creation of diverse and complicated game elements. However, it is not limited to video games, but also has applications in various other fields, including art and design, where the need for algorithmically generated content is widespread. Content in this context can therefore have a broad range of meanings: Images, 3D models, game levels, text, sound or a combination of these, to name just a few examples.\par 
The \gls{WFC} algorithm~\cite{gumin2016} is a a non-backtracking, greedy constraint solving method~\cite{karth2017,karth2022} that is able to generate complex patterns based on a set of input samples. It is known for its ability to create diverse and complex outputs that resemble the input samples while exhibiting novel combinations and variations. \Gls{WFC} employs two implementation strategies: the \emph{simple tiled model} and the \emph{overlapping model}, which share an identical algorithm core~\cite{nie2024}. In the simple tiled model, tilesets are manually prescribed with predefined adjacency constraints, whereas the overlapping model automatically generates this information from a sample input.\par
In both strategies, the output is compartmentalized into segments and the possibilities for each segment are iteratively constrained until a unique solution is determined. The term ``wave function collapse'' is borrowed from quantum physics because of the conceptual similarity. The ``wave function'' refers to the set of potential states of the segments, whereas the ''collapse`` occurs during the iterative process of narrowing down the possibilities.\par
Despite its name, \gls{WFC} is a purely classical algorithm. But even if \gls{WFC} has nothing to do with quantum physics beyond the terminology, the question arises as to whether quantum computers can still be used to execute a genuine quantum version of the algorithm. The motivation for leveraging quantum computers for \gls{PCG}, which is called \gls{QPCG} in the following, lies in their potential to introduce new levels of complexity, creativity, and efficiency. Because quantum physics exhibits intrinsic randomness~\cite{bera2017}, \glspl{QRNG} can be viewed as sources of true randomness~\cite{mannalatha2023}, making them a natural choice for generating randomized content. In addition, quantum superposition allows the random generation of complex patterns and variations that classical algorithms might find challenging to generate efficiently.\par
Currently available \gls{NISQ} hardware has very limited capabilities and is subject to significant noise and uncertainty~\cite{preskill2018}. While these kind of disruptive effects are typically to be avoided, they can also lead to novel and unexpected results for the generated content, potentially fostering creativity. In this sense, despite the technological limitations, it seems promising to use quantum computers already today for simple \gls{PCG} tasks and to study the results. For more complex tasks, hybrid quantum-classical algorithms are a promising strategy to overcome the limitations, for example by integrating quantum subroutines into classical \gls{PCG} methods to increase variability and complexity.\par
\Gls{QPCG} is therefore a promising approach, currently mainly for research, but in the future possibly also for practical applications. However, a general discussion of \gls{QPCG} is beyond the scope of this paper. Instead, the focus will specifically be on the \gls{WFC} algorithm and how it can be modified to enable the use of gate-based quantum computers~\cite{nielsen2010}. This paper has four main contributions:
\begin{itemize}
	\item A probabilistic formulation of the simple tiled model of \gls{WFC} is presented as a foundation for a quantum version. We denote this formulation as \gls{CWFC} to clearly distinguish between classical and non-classical approaches.
	\item The \gls{QWFC} method is proposed, which founds on the encoding of a probability distribution in a quantum circuit.	
	\item The \gls{HWFC} method is proposed to take into account the limitations of \gls{NISQ} hardware.
	\item The proposed methods are tested on simulators and \emph{IBM Quantum} devices.
\end{itemize}
The remainder of the manuscript is organized as follows. After a brief summary of related work, a formal description of \gls{CWFC} is given in an appropriate probabilistic form, which subsequently allows the development of \gls{QWFC} and \gls{HWFC}. The proposed methods are then demonstrated in practice. Finally, the paper ends with a conclusion.

\section{RELATED WORK}
There are many approaches to \gls{PCG}, such as machine learning~\cite{mao2024}, evolutionary algorithms~\cite{beukman2022} and search-based methods~\cite{zamorano2023}. \Gls{WFC} was originally proposed by Gumin~\cite{gumin2016} as an example of the latter. Subsequent research has extended the basic \gls{WFC} algorithm to address its limitations and expand its capabilities~\cite{karth2017,karth2022}. For example, by eliminating the need for two-dimensional grids~\cite{kim2019}, adding design constraints~\cite{sandhu2019}, or allowing interactive user control over the results~\cite{langendam2022,alaka2023}. Better scaling and runtime can be achieved by a hierarchical approach~\cite{beukman2023}, a nested approach~\cite{nie2024} or by using bitwise operations~\cite{punia2023}.\par
In contrast to classical \gls{PCG}, \gls{QPCG} is much less common in the literature. Previous work on \gls{QPCG} was mainly focusing on the quantum generalization of a blurring process~\cite{wootton2020a,wootton2023} and map generation using a quantum-enhanced decision making process~\cite{wootton2020b}. Moreover, related projects have been carried out in connection with the development of quantum-inspired games~\cite{piispanen2023}.

\section{CLASSICAL METHODS}
This section provides a formal framework for the following methods, referred to as \gls{PIPCG}. Subsequently, it is shown how the common \gls{CWFC} is a special case of \gls{PIPCG}. Only the main features of \gls{PIPCG} and \gls{CWFC} are summarized, the technical details can be found in the appendix.

\subsection{Probabilistic Iterative PCG (PIPCG)}
To begin with, it is first necessary to formally define the generated content at a sufficiently abstract level. For this purpose, it is presumed that any content instance $C$ can be described as an ordered sequence of $N$ segments
\begin{equation} \label{eqn:C}
	C := \{ (1,v_1), \dots, (N,v_N) \} \in \mathcal{C},
\end{equation}
where each segment $(i,v_i)$ is defined by a unique identifier $i \in [1,N]$ and a value $v_i \in A$ from an alphabet $A := [1,W]$ consisting of $W$ symbols with the notation $[i,j] := \{i,\dots,j\}$. The set of all possible content instances $\mathcal{C}$ contains $W^N$ elements. For example, if the content represents an image, its individual pixels can constitute the segments. Each pixel can attain one color from the available color space of the image (as its alphabet).

\subsubsection{Iterative Process.}
The procedural generation of content instance $C \in \mathcal{C}$ is realized through an iterative process, where each iteration $k \in [1,N]$ consists of three steps:
\begin{enumerate}
	\item Randomly select a new segment identifier $s(k) \in [1,N]$ from a set of available identifiers with a user-defined probability distribution.
	\item Randomly select a corresponding value $v_{s(k)} \in A$ from a set of available values  with a user-defined probability distribution.
	\item Add the chosen segment $(s(k), v_{s(k)})$ to the content instance.
\end{enumerate}
By repeating these steps, the content instance is assembled in the sense of $C(0) \overset{k=1}{\rightarrow} C(1) \overset{k=2}{\rightarrow} \cdots \overset{k=N}{\rightarrow} C(N)\equiv C$, where $C(k) := \{ (s(1),v_{s(1)}), \dots, (s(k),v_{s(k)}) \}$ represents the generated partial content instance at the end of iteration $k$. After $N$ iterations, the newly generated content instance is complete.

\subsubsection{Probabilistic Generation.}
An essential concept of \gls{PCG} is that only a limited subset of all possible content instances $C \in \mathcal{C}$ is generated according to the designer's aesthetic preferences, a given set of rules, or some other underlying logic; otherwise, the instances could be drawn directly from $\mathcal{C}$ with much less effort. For \gls{PIPCG}, the user-defined probability distributions for the segment identifiers and the corresponding values represent exactly this underlying logic.\par
Generating content with \gls{PIPCG} effectively corresponds to drawing a sample from the random variable $\rand{C} \sim p(C)$, where
\begin{equation} \label{eqn:pC}
	p(C) = p(v_1, \dots, v_N)
\end{equation}
denotes the probability to generate the content $C$ in the form of \cref{eqn:C} with values $v_1,\dots,v_N$. Hence, $\underline{\mathcal{C}} := \{ C \,|\, C \in \mathcal{C} \,\land\, p(C) > 0 \}$ represents the set of all content instances that can potentially be generated out of all possible content instances $\mathcal{C}$.

\subsection{Classical Wave Function Collapse (CWFC)}
\Gls{CWFC} can be understood as a special case of \gls{PIPCG}. In this manuscript, only a simplified form of the simple tiled model of \gls{CWFC} is considered (with the minor extension of graph-based adjacency~\cite{kim2019}), as it is sufficient to capture the key components of the approach. To realize \gls{CWFC}, the two randomized selection steps of \gls{PIPCG} are chosen as follows:
\begin{enumerate}
	\item For the segment identifier selection, the segments with the smallest Shannon entropy with respect to the available value options are chosen.
	\item Value selection is determined by a pattern-based set of user-defined rules.
\end{enumerate}

\subsubsection{Adjacency.}
\begin{figure}[!t]
	\centering
	\includegraphics[width=\linewidth]{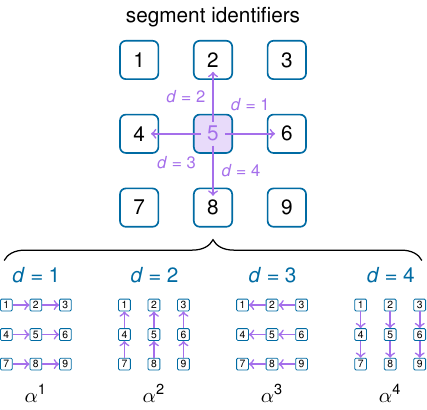}
	\caption{Exemplary nearest neighbor adjacency configuration for \gls{CWFC} on a two-dimensional grid consisting of $N=9$ segments. The adjacency relationship for each of the four directions \emph{right} ($d=1$), \emph{up} ($d=2$), \emph{left} ($d=3$), and \emph{down} ($d=4$) can be represented as a directed graph with adjacency matrix $\alpha^d$.}
	\label{fig:adjacency}
\end{figure}

For the pattern-based value selection, the segments are organized in such a way that they have a predefined adjacency relationship to each other that does not change. For example, consider that the pixels of an image constitute the segments. Choosing adjacency between the nearest neighbors leads to an adjacency configuration with four directions (right, up, left, down) as shown in \cref{fig:adjacency}.\par
Formally, each direction within a adjacency configuration is specified by an index $d \in [1,D]$, where $D$ denotes the total number of directions. The direction-based adjacency relationship between different segments is then defined by the coefficients $\alpha^d_{i,j} \in \{0,1\}$ for $i,j \in [1,N]$ and $d \in [1,D]$. If $\alpha^d_{i,j}=1$, segment $i$ is connected to segment $j$ in direction $d$; otherwise not. For each direction $d$, these coefficients form an adjacency matrix $\alpha^d \in \{0,1\}^{N \times N}$ of a directed graph with $N$ vertices that represent the segments. Hence, the segment adjacency is in fact an abstract concept that is not necessarily related to the visual form of the content.

\subsubsection{Patterns.}
Patterns define how values are chosen based on the already generated content. A pattern
\begin{equation} \label{eqn:P}
	P := \{ (d_1, v_1), \dots, (d_n, v_n) \} 
\end{equation}
consists of a set of $n \leq D$ direction-value pairs with $d_i \in [1,D]$ and $v_i \in A$ for $i \in [1,n]$, which represents a layout of segments. A pattern-based rule
\begin{equation} \label{eqn:rP}
	r_i^{\mathrm P}(v, P, u) = (v, u, P)
\end{equation}
is defined by a value $v \in A$, a factor $u \in \mathbb{R}_{>0}$, and a pattern $P$. It describes the weight $u$ of selecting a value $v$ for the target segment given the layout of adjacent segments $P$. The higher the value $u$, the more probable becomes the selection of the corresponding value. A selection mechanism with $m$ rules is then fully defined by a ruleset
\begin{equation} \label{eqn:R}
	R := \{r_1, \dots, r_m\}
\end{equation}
comprised of rules from \cref{eqn:rP} with $i\in[1,m]$.

\subsection{CWFC Example: Checkerboard}
As a simple example, consider the generation of images with black and white checkerboard patterns. The pixels of an image constitute the segments and the alphabet consists of only $W=2$ symbols standing for the two colors. As an adjacency configuration, the nearest-neighbor adjacency from \cref{fig:adjacency} is presumed.\par
\begin{figure}[!t]
	\centering
	\includegraphics[width=\linewidth]{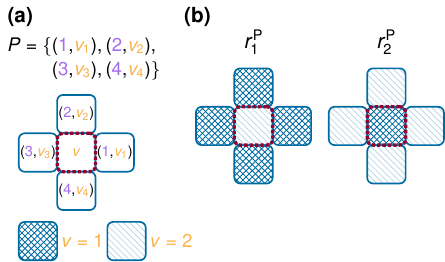}
	\caption{Visualization of pattern-based rules. (a) Visualization scheme. The center tile corresponds to the target segment $i$ with a target value of $v$, whereas the adjacent tiles represent the required values $v_1,\dots,v_4 \in \{1,2\}$ of the four adjacent segments in the respective directions $d=1,\dots,d=4$. (b) Pattern-based rules $r_1^{\mathrm P}(v=2, P=P_1, u=1)$ and $r_2^{\mathrm P}(v=1, P=P_2, u=1)$, \cref{eqn:rP}, that can be used for the generation of images with black and white checkerboard patterns.}
	\label{fig:patterns2d}
\end{figure}

A checkerboard image can be achieved with only two patterns of the form of \cref{eqn:P}, namely $P_1 := \{ (1, 1), (2, 1), (3, 1), (4, 1) \}$ and $P_2 := \{ (1, 2), (2, 2), (3, 2), (4, 2) \}$. The corresponding ruleset, \cref{eqn:R}, reads $R = \{ r_1^{\mathrm P}, r_2^{\mathrm P} \}$ and therefore contains the $m=2$ rules $r_1^{\mathrm P}(v=2, P=P_1, u=1)$ and $r_2^{\mathrm P}(v=1, P=P_1, u=1)$, see \cref{fig:patterns2d}.\par
\begin{figure}[!t]
	\centering
	\includegraphics[width=\linewidth]{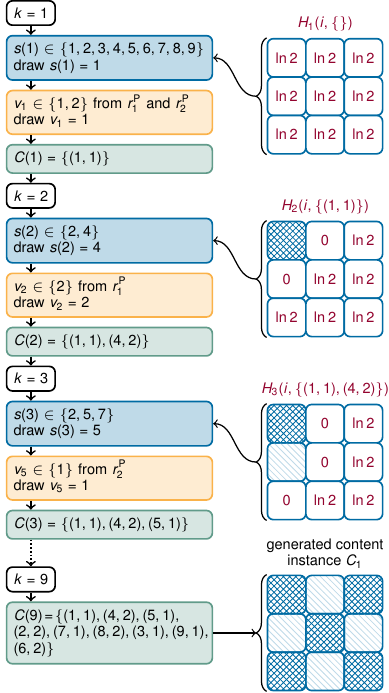}
	\caption{\gls{CWFC} for the generation of $3 \times 3$ images with checkerboard patterns defined by the rules from \cref{fig:patterns2d}. In each iteration $k \in [1,9]$, three steps take place: (i) a segment identifier $s(k)$ with the smallest entropy is drawn, (ii) a corresponding value $v_{s(k)}$ is drawn according to the pattern-based ruleset, and (iii) the newly generated identifier-value pair is added as a segment to the content instance $C(k)$. For $s(k)$ and $v(k)$, the uniformly distributed set of possible choices are listed. The choices for $s(k)$ depend on the Shannon entropy $H_k$ of each undefined segment (see appendix), as shown on the right. The choices for $v(k)$ depend on the fulfillment of the two rules $r_1^{\mathrm P}$ and $r_2^{\mathrm P}$, as listed.}
	\label{fig:cgeneration2d}
\end{figure}

The iterative process of \gls{CWFC} is sketched in \cref{fig:cgeneration2d} for a $3 \times 3$ image (represented by $N=9$ segments). Only two content instances can be generated, $C_1 := \{ (1, 1), \dots \}$ and $C_2 := \{ (1, 2), \dots \}$ with equal probability $p(C_1) = p(C_2) = \frac{1}{2}$, \cref{eqn:pC}. The choice of the first value $v_{s(1)}$ already determines the choices of all other values.

\section{QUANTUM METHODS}
In the previous section, only classical \gls{PCG} was considered. This section is about \gls{QPCG}. First, \gls{QWFC} is proposed as a special case of \gls{PIPCG}. For this purpose, the probability distribution of the content instances $p(C)$, \cref{eqn:pC}, is realized with the help of a quantum circuit and the sampling is performed by exploiting the intrinsic randomness of quantum measurements. That is, the \emph{actual} quantum mechanical wave function collapse is used to implement a quantum version of the Wave Function Collapse algorithm.\par
Subsequently, this section proposes \gls{HWFC} as a quantum-classical hybrid algorithm based on content partitioning to overcome the limitations of \gls{NISQ} devices. Only the main features of \gls{QWFC} and \gls{HWFC} are summarized, the technical details can be found in the appendix. A software implementation is available online~\cite{heese2023}.

\subsection{Quantum Wave Function Collapse (QWFC)}
To realize \gls{QWFC}, the two randomized selection steps of \gls{PIPCG} are chosen as follows:
\begin{enumerate}
	\item The segment identifier selection is performed with a predefined order given by the vector $\sigma \in S_N$, where $S_N$ is the set of all permutations of $[1,N]$.
	\item The value selection is based on a pattern-based ruleset in the same way as for \gls{CWFC}.
\end{enumerate}

\subsubsection{Value Encoding.}
The value for each segment is encoded by a set of qubits, which requires $q := \lceil \log_2 W \rceil$ qubits for each segment $i \in [1,N]$. The group of $q$ qubits that represent the value for segment $i$ is denoted by $\mathcal{Q}_i := \{ q_i^1, \dots, q_i^q \}$, where $q_i^j$ references the $j$th qubit of this group for $j \in [1,q]$. The state of these qubits encodes the value $v_i$ of the segment $i$ in binary as the tensor product sequence of qubit states
$\bigotimes_{j=1}^q \ket{b_j} = \ket{v_i-1}_i \in \mathcal{H}_i$ via $v_i-1 = \sum_{j=1}^{q} b_j 2^{j-1}$ with $b_j \in \{0,1\}$. Here, $\mathcal{H}_i$ denotes the joint Hilbert space of all qubit from $\mathcal{Q}_i$ with $v_i \in A$ for $i \in [1,N]$.\par
In total, $Q := N q$ qubits are required to represent the value of all $N$ segments. The joint Hilbert space of all qubits from $\bigcup_{i=1}^N \mathcal{Q}_i$ is given by $\mathcal{H} := \bigotimes_{i=1}^N \mathcal{H}_i$.

\subsubsection{Probability Encoding.}
\begin{figure}[!t]
	\centering
	\includegraphics[width=\linewidth]{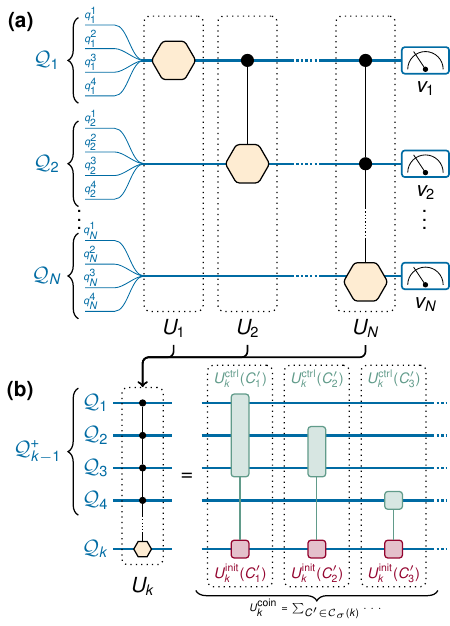}
	\caption{\gls{QWFC} circuit with $\sigma = (1,\dots,N)$. (a) Circuit layout with an exemplary alphabet $A$ with $W=16$ symbols, which requires a set $\mathcal{Q}_i$ of $q=4$ qubits for each segment with $i \in [1,N]$. In each iteration $k \in [1,N]$, the operator $\operator{U}_k$ prepares the qubits from $\mathcal{Q}_k$ in a superposition state $\ket{C'}_{k}$, entangled with (a subset of) the qubits from $\mathcal{Q}_1,\dots,\mathcal{Q}_{k-1}$. All qubits are measured to obtain $p(C)$ (see appendix). (b) Gate decomposition of $\operator{U}_k$ into the components of $\operator{U}_k^{\mathrm{coin}}$, which consists of pairs $\operator{U}_k^{\mathrm{ctrl}}(C') \otimes \operator{U}_k^{\mathrm{init}}(C')$. Each of these pairs represent a conditional loading of a probability distribution. The combined control operators act on the qubits from $\mathcal{Q}_k^+$, whereas the initialization operators act on the qubits from $\mathcal{Q}_k$ (see appendix).}
	\label{fig:qwfccircuit}
\end{figure}

Presume a state $\ket{\Psi} \in \mathcal{H}$ that can realize samples from $p(C)$, \cref{eqn:pC}, with
\begin{equation} \label{eqn:pCQ}
	p(C\!=\!\{(1,v_1),\dots,(N,v_N)\}) = \big\vert \bra{\Psi} \bigotimes_{i=1}^{N} \ket{v_i - 1}_i \big\vert^2.
\end{equation}
To prepare $\ket{\Psi}$ from a ground state $\ket{0} := \bigotimes_{i=1}^{N} \ket{0}_i$ based on a pattern-based ruleset $R$, \cref{eqn:R}, a series of operators is applied, one for each iteration $k\in[1,N]$ such that $\ket{\Psi} := \operator{U}_N \cdots \operator{U}_1 \ket{0}$ represents the final state after $k$ iterations. In each iteration $k$, the state of the qubits $\mathcal{Q}_{\sigma_k}$ (that represent the value for segment $\sigma_k$) is prepared conditioned on the state of (a subset of) the qubits $\mathcal{Q}_{\sigma_1},\dots,\mathcal{Q}_{\sigma_{k-1}}$ (that represent the values of the already prepared segments $\sigma_1,\dots,\sigma_{k-1}$), which leads to an entangled joint state as sketched in \cref{fig:qwfccircuit}(a).\par
The required number of qubits for the proposed \gls{QWFC} method increases linearly with the number of segments and logarithmically with the number of symbols in the alphabet. At the same time, the respective circuits may become exponentially deeper because of the costly conditional preparation of probability distributions~\cite{dasgupta2022}.

\subsection{QWFC Example: Checkerboard}
\begin{figure}[!t]
	\centering
	\includegraphics[width=\linewidth]{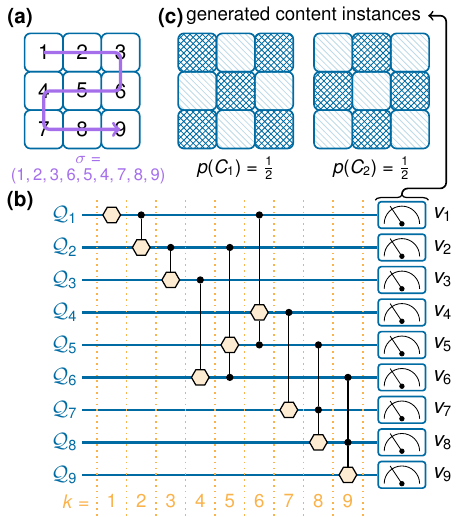}
	\caption{\gls{QWFC} for the generation of $3 \times 3$ images with black and white checkerboard patterns in analogy to \cref{fig:cgeneration2d}. (a) Predefined segment order $\sigma = (1,2,3,6,5,4,7,8,9)$. (b) Circuit layout with the same symbols as in \cref{fig:qwfccircuit}. (c) Generated content instances $C_1$ and $C_2$ with $p(C_1)=p(C_2)=\frac{1}{2}$.}
	\label{fig:qgeneration2d}
\end{figure}

In the following, the above example for the generation of $3 \times 3$ checkerboard images is considered for \gls{QWFC}, where the predefined segment order $\sigma = (1,2,3,6,5,4,7,8,9)$ is used as visualized in \cref{fig:qgeneration2d}(a). The resulting \gls{CWFC} circuit is shown in \cref{fig:qgeneration2d}(b) and the resulting probability distribution of content instances $p(C)$, \cref{eqn:pCQ}, in \cref{fig:qgeneration2d}(c). A (noise-free) measurement of the circuit corresponds to drawing a sample from this distribution.

\subsection{Hybrid Quantum-Classical WFC (HWFC)}
\begin{figure}[!t]
	\centering
	\includegraphics[width=\linewidth]{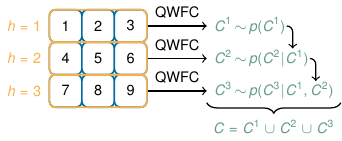}
	\caption{\gls{HWFC} with $H=3$ partitions of $N=9$ segments. For each partition, a \emph{conditional} \gls{QWFC} is performed and the resulting content instances are combined.}
	\label{fig:hybridwfc}
\end{figure}

\Gls{HWFC} represents a resource-efficient alternative to \gls{QWFC} that might be more suitable for \gls{NISQ} devices. The key idea of this approach is to separate the content into $H$ partitions, each containing a subset of the segments as sketched in \cref{fig:hybridwfc}. For each partition $h \in [1,H]$, a \gls{QWFC} is performed, which yields the content instance partition $C^h$ as a sample of the random variable $\rand{C}^h \sim p(C^h) := p^h(C^h | C^1, \dots, C^{h-1})$ in analogy to \cref{eqn:pCQ}. To apply \gls{QWFC} on the partition, two modifications are required. First, the iteration over all segment identifiers $[1,N]$ is replaced by the iteration over the segment identifiers from $S^h$. Second, the already sampled sub-contents $C^1,\dots,C^{h-1}$ from the partitions $[1,h-1]$ are used as an additional constraint for the pattern-based rules. This modification induces classical correlations from previously sampled content partitions into the circuit and therefore requires a \emph{conditional} form of \gls{QWFC} that respects these correlations as constraints. The resulting distribution of the joint content instances is then given by $p\left(C = \bigcup_{h=1}^H C^h \right) = \prod_{h=1}^H p^h(C^h | C^1, \dots, C^{h-1})$
and requires less quantum resources than performing a single run of \gls{QWFC} on all segments.

\section{DEMONSTRATION}
In the present section, five simple content creation use cases are presented to demonstrate the proposed methods. For \gls{QWFC} and \gls{HWFC}, both idealized (\ie, noise-free) simulators and \emph{IBM Quantum} devices were used, the latter being accessed via the \emph{IBM Quantum Cloud Services}~\cite{ibmq2021} during December 2023.

\subsection{Checkerboard Revisited}
\begin{figure}[!t]
	\centering
	\includegraphics[width=\linewidth]{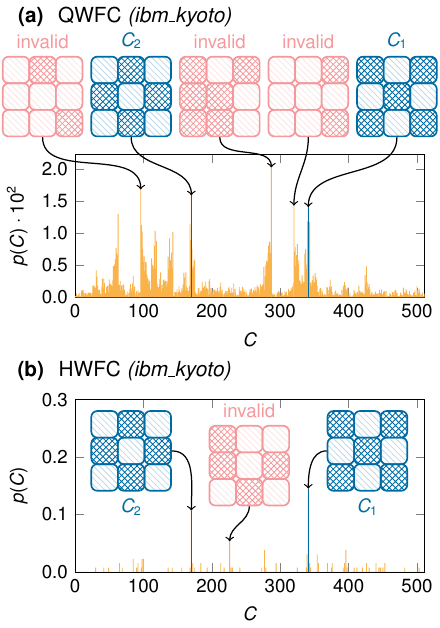}
	\caption{Checkerboard generation. Plots show the generated content instances $C$, \cref{eqn:C}, (encoded as integers in $[0,511]$ from the measured bit strings) and their respective probabilities $p(C)$, \cref{eqn:pCQ}. (a) \gls{QWFC} on the \emph{IBM Quantum} device \emph{ibm\_kyoto}. The dominating instances ($p(C) > \num{1.4e-2}$) are highlighted. (b) \gls{HWFC} with $H=3$ on the same device results in less invalid instances.}
	\label{fig:chess}
\end{figure}

The first use case is the previously presented checkerboard example, \cf \cref{fig:patterns2d,fig:qgeneration2d}. Again, $3 \times 3$ instances are considered, for which a comparison between \gls{QWFC} and \gls{HWFC} is performed on the \emph{IBM Quantum} device \emph{ibm\_kyoto} (\kyoto). For \gls{QWFC}, \num{10000} shots are measured, each yielding a content instance. For \gls{HWFC}, $H=3$ equal-sized partitions are used in analogy to \cref{fig:hybridwfc} and the method is repeated \num{131} times, each time yielding a content instance. The results are shown in \cref{fig:chess}.\par
In total, there are $2^9=512$ possible content instances $C \in \mathcal{C}$, but only two of them, $C_1$ and $C_2$, are valid. The remaining \num{510} content instances are invalid because they violate the prescribed patterns. In \cref{fig:chess}(a), \gls{QWFC} is able to generate the two valid instances, but at the same time also produces a lot of invalid instances due to hardware imperfections. From \cref{fig:chess}(b) it can be seen that the results from \gls{HWFC} are much closer to the idealized solution that would contain only $C_1$ and $C_2$ with equal probability. It is a consequence of the reduced circuit size from this method that invalid solutions occur with much less frequency than for \gls{QWFC}.

\subsection{Pipes}
\begin{figure}[!t]
	\centering
	\includegraphics[width=\linewidth]{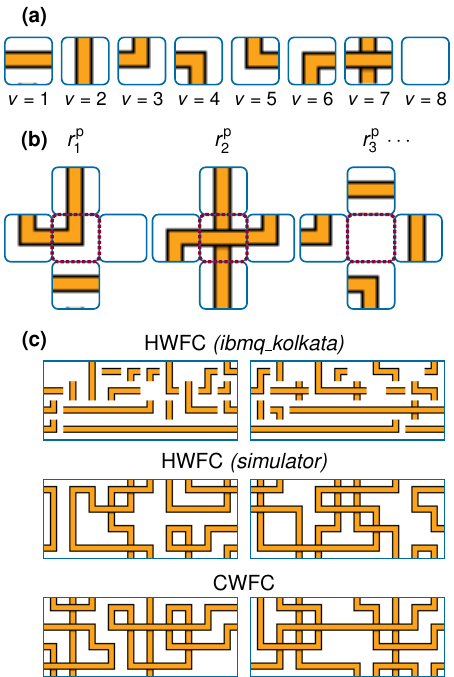}
	\caption{Pipe generation. The generated image is required to show a network of connected pipes. (a) Alphabet consisting of $W=8$ symbols, each of which represents a tile with a pipe section and an empty tile. (b) Example of three pattern-based rules $r_1^{\mathrm p}$, $r_2^{\mathrm p}$, and $r_3^{\mathrm p}$, \cf \cref{fig:patterns2d}. (c) Example $10 \times 4$ tile images that were generated using \gls{HWFC} with $H=10$ on the \emph{IBM Quantum} device \emph{ibm\_kolkata} and a simulator; also shown are images from \gls{CWFC}.}
	\label{fig:pipes}
\end{figure}

The second use case is to create an image from tiles of seven different pipe (or line) sections as well as one blank tile. That is, the alphabet consists of $W=8$ symbols and each tile in the image constitutes a segment. The alphabet is shown in \cref{fig:pipes}(a). As adjacency configuration, the nearest-neighbor adjacency from \cref{fig:adjacency} is presumed with $D=4$. The pattern-based ruleset is chosen in such a way that neighboring tiles have to form a connecting network of pipes. Three example rules for this purpose are shown in \cref{fig:pipes}(b). In total, $m=2048$ rules are required to achieve this goal.\par 
Exemplarily, images consisting of $10 \times 4$ tiles ($N=40$) are considered, for which a comparison between \gls{HWFC} on the \emph{IBM Quantum} device \emph{ibm\_kolkata} (\kolkata), \gls{HWFC} on a simulator and \gls{CWFC} is performed. For \gls{HWFC}, $H=10$ equal-sized partitions are used, each corresponding to one column of the image. A few resulting instances are shown in \cref{fig:pipes}(c).\par
Based on a visual comparison, both \gls{CWFC} and \gls{HWFC} on the simulator yield similar results, as expected. However, the results from \emph{ibm\_kolkata} clearly contain a lot of invalid patterns as a result of the imperfect hardware. This demonstrates the limitations of the proposed method on \gls{NISQ} devices.

\subsection{Hexagon Map}
\begin{figure}[!t]
	\centering
	\includegraphics[width=\linewidth]{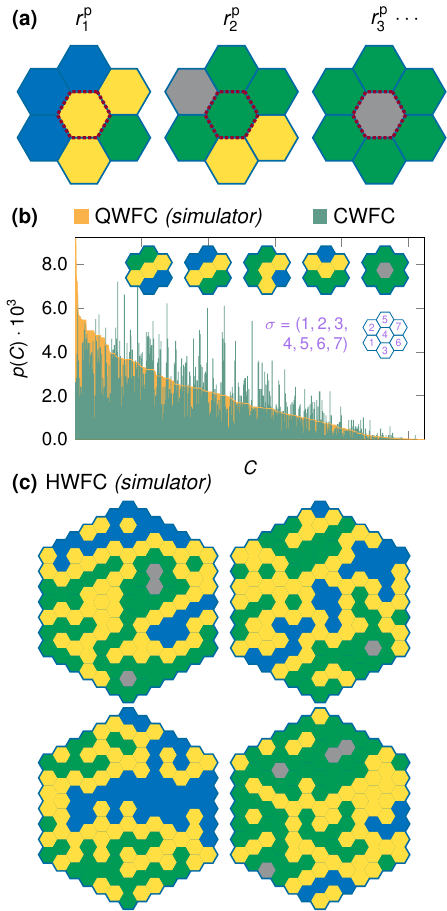}
	\caption{Map generation. Only connections of the form blue-yellow-green-gray are allowed. (a) Example of three pattern-based rules $r_1^{\mathrm p}$, $r_2^{\mathrm p}$, and $r_3^{\mathrm p}$. (b) Distribution of content instances $p(C)$ for $N=7$ hexagons using \gls{QWFC} on a simulator (sorted by descending probability) and \gls{CWFC} (with the same ordering). The inset plot shows the predefined segment order $\sigma$ and the five most probable instances for \gls{QWFC}. The asymmetry is a consequence of the choice of $\sigma$. (c) Example images that were generated from $N=127$ hexagons using \gls{HWFC} with $H=20$ on a simulator.}
	\label{fig:hexagons}
\end{figure}

The third use case addresses images based on hexagonal tiles that can be interpreted as a map consisting of different terrains. For this purpose, an alphabet of $W=4$ symbols is chosen, each corresponding to a unicolored tile (blue, yellow, green, and gray) and each tile constitutes a segment. As adjacency configuration, a nearest-neighbor adjacency is presumed ($D=6$). The prescribed pattern-based ruleset only allows connections of the form blue-yellow-green-gray, which can be fulfilled with $m=1586$ rules, see \cref{fig:hexagons}(a). To promote a higher occurrence of blue tiles, the factors in \cref{eqn:rP} are chosen as $u=5$ for blue tiles and $u=1$ otherwise.\par
As a first example, images consisting of $N=7$ tiles are considered, for which \gls{QWFC} on a simulator and \gls{CWFC} are performed. The resulting distributions of content instances $p(C)$ in \cref{fig:hexagons}(b) is exactly calculated for \gls{QWFC}, \cref{eqn:pCQ}, and based on \num{10000} samples for \gls{CWFC}, \cref{eqn:pC}. Both distributions are significantly different, which is no surprise since the two approaches use a different segment identifier selection. The second example considers images from $N=127$ tiles, for which \gls{HWFC} with $H=20$ is performed on a simulator. Example instances are shown in \cref{fig:hexagons}(c). 

\subsection{Platformer}
\begin{figure}[!t]
	\centering
	\includegraphics[width=\linewidth]{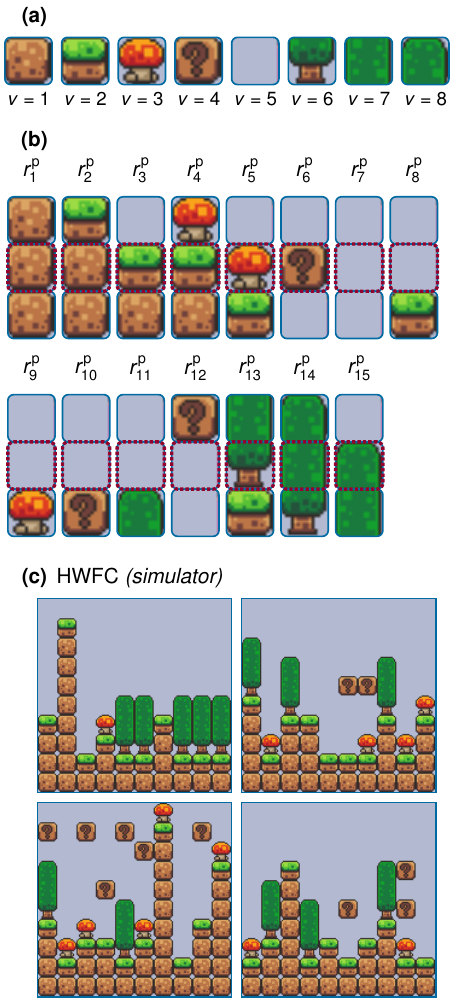}
	\caption{Level generation. (a) Alphabet consisting of $W=8$ symbols, each of which represents a tile~\cite{platformer2019}. (b) Pattern-based ruleset consisting of $m=15$ rules. (c) Example $10 \times 10$ tile images that were generated using \gls{HWFC} with $H=20$ on a simulator.}
	\label{fig:platformer}
\end{figure}

The fourth use case is motivated by computer game level design. The goal is to create a platformer-type level from a set of eight tiles~\cite{platformer2019}, each corresponding to a different game element (ground, grass, mushroom, block, air, and a tree that is composed of three tiles). Consequently, the alphabet consists of $W=8$ symbols, see \cref{fig:platformer}(a). As adjacency configuration, the two tiles above and below are taken into account as neighbors ($D=2$). A pattern-based ruleset is prescribed that meets five requirements: (i) The bottom row consists of ground tiles and ground tiles can only be placed on top of each other, (ii) only a grass tile or a mushroom tile can be placed above a ground tile, (iii) tree tiles must be placed in order with the lowest tile above a grass tile, (iv) air tiles can only be placed above grass, mushroom or other air tiles, (v) block tiles can only be placed between two air tiles. These requirements can be fulfilled with $m=15$ rules, see \cref{fig:platformer}(b). The first requirement can be resolved with a functional factor $u(i)$ in \cref{eqn:rP} that vanishes for all non-ground tiles in the bottom row (see appendix). For all non-vanishing cases, $u=\num{0.1}$ for $r_6^{\mathrm p}$ and $u=1$ otherwise to reduce the occurrence of block tiles. The segment order $\sigma$ is chosen such that the levels are built up from bottom to top.\par
Exemplarily, levels with $10 \times 10$ tiles are considered, for which \gls{HWFC} with $H=20$ equal-sized partitions (each corresponding to a half row of the level) is performed on a simulator. Four resulting instances are shown in \cref{fig:platformer}(c).

\subsection{Voxel Skyline}
\begin{figure}[!t]
	\centering
	\includegraphics[width=\linewidth]{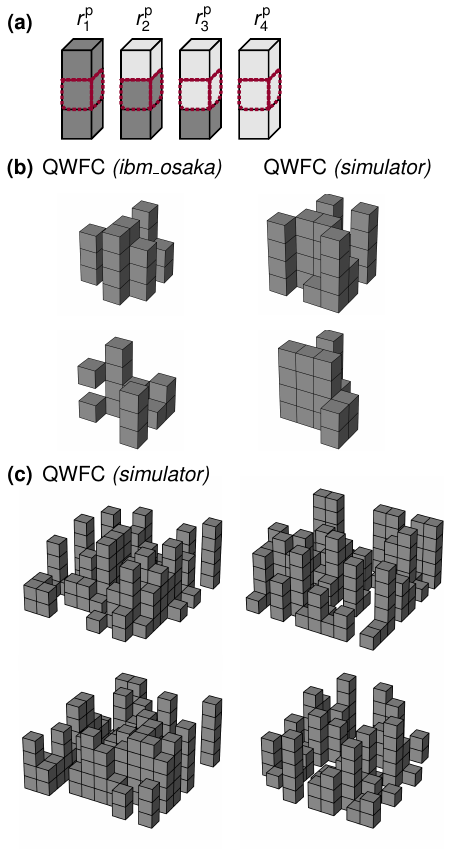}
	\caption{Skyline generation. (a) Pattern-based ruleset consisting of $m=4$ rules in an analogous representation as in the previous figures. (b) Example images from a $4 \times 4 \times 4$ voxel grid that were generated using \gls{QWFC} on the \emph{IBM Quantum} device \emph{ibm\_osaka} and a simulator. (c) Example images from a $10 \times 10 \times 5$ voxel grid that were generated using \gls{QWFC} on a simulator.}
	\label{fig:voxels}
\end{figure}

The final use case considers the creation of a three-dimensional voxel graphic with a binary alphabet of $W=2$ symbols, which represent the presence or absence of a voxel. Each voxel constitutes a segment, as adjacency configuration the voxel above and below are taken into account ($D=2$). The chosen pattern-based ruleset with $m=4$ ensures that voxels are built from the ground up, see \cref{fig:voxels}(a). This leads to a skyline-like structure.
Exemplarily, two cases are considered: First, images from a $4 \times 4 \times 4$ voxel grid, for which \gls{QWFC} is performed both on a the \emph{IBM Quantum} device \emph{ibm\_osaka} (\osaka) and a simulator, see \cref{fig:voxels}(b). An error violating $r_3^{\mathrm p}$ can be seen in the bottom left image that results from hardware imperfections. Second, images from a $10 \times 10 \times 5$ voxel grid, for which \gls{QWFC} is performed on a simulator, see \cref{fig:voxels}(c).

\section{CONCLUSION}
In this work, \gls{QWFC} was proposed as a method to realize \gls{WFC} on a gate-based quantum computer as an example of \gls{QPCG}. The idea is to construct a quantum circuit from which valid content instances (that fulfill the prescribed patterns) can be sampled via measurements. In other words, the quantum circuit acts as a special-purpose \gls{QRNG} for content of a desired form that makes use of the intrinsic randomness of quantum physics. This means that the actual quantum mechanical collapse of the wave function is used for the implementation, from which the classical method derived its name.\par
The proposed method is not without its challenges. The generating quantum circuit becomes larger from both the content and alphabet size and at the same time deeper from correlations within the content instance distribution, which makes it difficult to evaluate practical use cases on current hardware. For this reason, \gls{HWFC} was proposed as a hybrid method that partitions the generation into smaller tasks for \gls{QWFC}.\par
The experimental results have shown the limitations of \gls{QWFC} and \gls{HWFC}, but have also proven that at least simple examples can already be implemented on today's quantum hardware. The biggest bottleneck of the circuit complexity is the conditional preparation of probability distributions. Investigating possible improvements in this area with regard to specific \gls{QPCG} use cases could therefore be a potential direction for future research.\par
The conceptual advantage of \gls{QWFC} is that a circuit, once designed, represents the distribution of content instances such that new instances can be generated by simple measurements without any additional algorithmic effort. A shortcoming of \gls{QWFC} is that only a fixed order of segments is considered instead of a dynamic (\eg, entropy-based) order that depends on the previously selected segments, which is a  chance for future improvements.\par
In conclusion, the paper aims to provide an initial approach to \gls{QPCG} that demonstrates its feasibility but leaves room for improvement and raises the question of useful application areas given the limits of \gls{NISQ} devices.

	\section{APPENDIX}
This appendix section contains the technical details of the presented \gls{PCG} methods.

\subsection{Technical Details of PIPCG}
Presume the generated partial content instance $C(k) \in \mathcal{C}(k)$ for $k \in [1,N]$, where $\mathcal{C}(k)$ denotes all possible partial content instances at the end of iteration $k$ with $\mathcal{C}(N) \equiv \mathcal{C}$. One has $\vert C(k) \vert = k$, $C(k) = C(k-1) \cup \{ (s(k),v_{s(k)})\}$, $s(i) \neq s(j) \,\forall\, i, j \in [1,k] \land i \neq j$, and $\vert \mathcal{C}(k) \vert = \sum_{I \in J(k)} W^{|I|}$, where $J(k) := \{ I \,|\, I \in [1,N] \land \vert I \vert = k \}$ with $\vert J(k) \vert = {N \choose k}$. Let $C(0) := \{\}$ and $\mathcal{C}(0) := \{C(0)\}$. Furthermore, $S^+(C(l)) := \{s(1), \dots, s(l)\}$ with $\vert S^+(C(l)) \vert=l$ and $S^-(C(l)) := [1,N] \setminus S^+(C(l))$ with $\vert S^-(C(l)) \vert=N-l$ for all $C(l) \in \mathcal{C}(l)$ and for all $l \in [0,N]$.\par
In the following, the three steps within each iteration $k$ of \gls{PIPCG} are explained. In the first step, an identifier is selected by drawing a sample $s(k)$ from the random variable
\begin{equation} \label{eqn:s}
	\rand{s}(k) \sim p^{\mathrm s}_k(i=s(k)|C'=C(k-1)).
\end{equation}
The probability distribution $p_k^{\mathrm s}(i | C')$ with support $i \in [1,N]$ and $C' \in \mathcal{C}(k-1)$ is a user-defined parameter of the procedure that determines how new segment indices $k$ are selected based on the partial content instance $C'=C(k-1)$ that has already been generated up to the previous iteration $k-1$ (making \gls{PIPCG} non-Markovian). Two conditions must hold:
\begin{Scondenum}
	\item \label{ps1} No identifier can be chosen twice, \ie, $p_l^{\mathrm s}(i=s | C'=C(l-1))=0 \,\forall\, s \in S^+(C(l-1)) \,\forall\, C(l-1) \in \mathcal{C}(l-1) \,\forall\, l \in [1,N]$.
	\item \label{ps2} At least one valid identifier has to be available, \ie, $\forall\, l \in [1,N] \,\forall\, C(l-1) \in \mathcal{C}(l-1) \,\exists\, s \in S^-(C(l-1)) : p_l^{\mathrm s}(i=s | C'=C(l-1)) > 0$.
\end{Scondenum}\par
In the second step, a value $v_{s(k)}$ is selected for the segment by drawing a sample $v_{s(k)}$ from the random variable
\begin{equation} \label{eqn:v}
	\rand{v}_{s(k)} \sim p^{\mathrm v}_k( \nu=v_{s(k)} | i=s(k), C'=C(k-1) ).
\end{equation}
The probability distribution $p^{\mathrm v}_k( \nu | i, C' )$ with support $\nu \in A$, $i \in [1,N]$ and $C' \in \mathcal{C}(k-1)$ is a user-defined parameter of the procedure that determines how new values are selected based on the selected identifier $i=s(k)$ of the current iteration and $C'=C(k-1)$. Two conditions must hold:
\begin{Vcondenum}
	\item \label{pv1} Only values from the alphabet are available, \ie, $p^{\mathrm v}_l( \nu=v | i=s, C'=C(l) )=0 \,\forall\, v \not\in A \,\forall\, s \in [1,N] \,\forall\, C(l) \in \mathcal{C}(l-1) \,\forall\, l \in [1,N]$.
	\item \label{pv2} At least one value from the alphabet has to be available, \ie, $\forall\, l \in [1,N] \,\forall\, C(l-1) \in \mathcal{C}(l-1) \forall s \in S^-(C(l-1)) \exists v \in A : p_l^{\mathrm v}( \nu=v | i=s, C'=C(l) ) > 0$.
\end{Vcondenum}\par
In the third step, $C(k) = C(k-1) \cup \{ (s(k),v_{s(k)})\}$. The process is repeated for $k \rightarrow k+1$ until $k=N$.\par
The probability to generate $C(k)$ is given by $p(C(k)) := \sum_{\sigma \in S_k(C(k))} \prod_{l=1}^k q_l^k(\sigma, C(k))$, where
\begin{align} \label{eqn:qlk}
	q_l^k(\sigma, C(k)) :=  p_l(& i=\sigma_l, \nu=v_{\sigma_l} | \nonumber \\
	& C' \!=\! \{ (\sigma_1, v_{\sigma_1}), \!\dots,\! (\sigma_{l-1}, v_{\sigma_{l-1}}) \} ).
\end{align}
with $p_l( i, \nu | C') := p_k^{\mathrm s}( i | C') p_k^{\mathrm v}( \nu | i, C')$, $S_k(C(k))$ as the set of all permutations of $S^+(C(k))$ with $\vert S_k(C(k)) \vert = k!$, and the $l$th value of a permutation $\sigma \in S_k(C(k))$ denoted by $\sigma_l$ for $l \in [1,k]$. In \cref{eqn:qlk}, let $C'=\{\}$ for $l=1$. $v_{\sigma_m}$ represents the value of the segment with identifier $\sigma_m$ from the content instance $C(k)$ for $m \in [1,l]$. The support $\underline{\mathcal{C}}(k) := \{ C(k) \,|\, C(k) \in \mathcal{C}(k) \land p(C(k)) > 0 \}$ of $p(C(k))$ contains all $C(k)$ that can be generated for the given $k$. For $k=N$, \cref{eqn:pC} emerges with $p(v_1, \dots, v_N) := \sum_{\sigma \in S_N} \prod_{l=1}^N q_l^N(\sigma, C)$ and $\underline{\mathcal{C}}(N) = \underline{\mathcal{C}}$.

\subsection{Technical Details of CWFC}
For the value selection, a general rule-based selection mechanism is introduced first and then the pattern-based selection is presented as a special case. Consider a set of $m$ rules $R \in \mathcal{R}$, \cref{eqn:R}, with the set of of all possible rulesets $\mathcal{R}$. Each rule $r_i := r_i(v,w) := (v, w)$ for $i \in [1,m]$ consists of a value $v \in A$ and a weight function $w := w(i, \alpha^d, C') \in \mathbb{R}_{\geq 0}$ with $w : [1,N] \times \mathcal{B} \times \overline{\mathcal{C}} \rightarrow \mathbb{R}_{\geq 0}$, where $\overline{\mathcal{C}} := \cup_{k=0}^{N-1} \mathcal{C}(k)$ and $\mathcal{B} := \{0,1\}^{N \times N}$. Given $R \in \mathcal{R}$, $p_k^{\mathrm v}(\nu | i, C') = p_k^{\mathrm{v,R}}(\nu | i, C') := \frac{F_k(\nu, i, C')}{\overline{F}_k(i, C')}$, \cref{eqn:v}, with $F_k(\nu, i, C') := \sum_{\substack{(v, w) \in R \\ v=\nu}} w(i, \alpha^d, C')$ and $\overline{F}_k(i, C') := \sum_{\nu \in A} F_k(\nu, i, C')$ for $\nu \in A$, $i \in [1,N]$, $C' \in \mathcal{C}(k)$ for $k \in [1,N]$, and $\alpha^d \in \mathcal{B}$ for $d \in [1,D]$. \Cref{pv1} is always satisfied, whereas in case of $\overline{F}_k(i, C') \overset{!}{=} 0$ for any $C' \in \mathcal{C}(k)$ with $p_k^{\mathrm{s,H}}(i | C') p(C') > 0$, \cref{pv2} is violated.\par
Pattern-based value selection is a special case of rule-based value selection, where $w(i, \alpha^d, C') \!:=\! w^{\mathrm P}(i, \alpha^d, C', P, u) \!:=\! u \tau (i, \alpha^d, C', P)$ of each rule $r_i := r_i^{\mathrm P}(v, P, u) := ( v, w^{\mathrm P}(i, \alpha^d, C', P, u) )$ for $i \in [1,m]$ is defined by $u \in \mathbb{R}_{>0}$ and $P \in \mathcal{P}$, \cref{eqn:P} with $d_i \in [1,D]$ for $i \in [1,n]$, $d_i \neq d_j \,\forall\, i, j \in [1,n] \land i \neq j$, and $v_i \in A$ for $i \in [1,n]$. The set of all possible patterns reads $\mathcal{P} := \{P \,|\, P \in ([1,D] \times A)^n \,\forall\, n \in [0,D] \}$. A pattern $P$ is fulfilled, if $\tau (i, \alpha^d, C', P) := \prod_{(d,v) \in P} \prod_{(s,v') \in C' \land \alpha^d_{i,s}=1} \delta_{v v'} \in [0,1]$ evaluates to one. The set of all possible pattern-based rulesets is denoted by $\mathcal{R}^{\mathrm P} \subset \mathcal{R}$. Summarized, the weight function $w(i, \alpha^d, C')$, yields $u$ if the pattern $P$ applies and zero otherwise. Only adjacent segments (\ie, $\alpha^d_{i,s}=1$) have an influence on $p_k^{\mathrm v}(\nu | i, C')$.\par 
The rules $r_i$ can be rewritten in form of \cref{eqn:rP}. One has $p_k^{\mathrm{v,R}}(\nu | i, C') = p_k^{\mathrm{v,P}}(\nu | i, C') := \sum_{\substack{(v, u, P) \in R \\ v=\nu}} \gamma(u, i, C') \tau (i, \alpha^d, C', P)$ with $\gamma(u, i, C') := \frac{u}{\overline{F}_k(i, C')} \in [0,1]$. Optionally, a functional factor instead of a constant factor can be used, \ie, $u := u(i,C) : [1,N] \times \overline{\mathcal{C}} \rightarrow \mathbb{R}$.\par
For the entropic segment identifier selection, one has
\begin{equation} \label{eqn:psH}
	p_k^{\mathrm s}(i | C') = p_k^{\mathrm{s,H}}(i | C') := \begin{cases} \vert I_k(C') \vert^{-1}  & \text{if } i \in I_k(C') \\ 0 & \text{otherwise}\end{cases}
\end{equation}
in \cref{eqn:s} with $I_k(C') := \{ i \,|\, i \in S^-(C') \land H_k(i,C') = h_k(C') \}$ and $h_k(C') := \argmin_{j \in S^-(C')} H_k(j,C')$, where
$H_k(i,C') := - \sum_{v \in A} p_k^{\mathrm{v,P}}( \nu=v | i, C') \ln p_k^{\mathrm{v,P}}( \nu=v | i, C')$ denotes the Shannon information entropy for $C' \in \mathcal{C}(k-1)$, $k \in [1,N]$.

\subsection{Technical Details of QWFC}
Presume a predefined order of segment indices $\sigma \in S_N$. Hence, $p_k^{\mathrm s}(i | C') = p_k^{\mathrm{s,o}}(i | C') := \delta_{i \sigma_i}$ and $\mathcal{C}(k) = \mathcal{C}_{\sigma}(k) := \{ C \,|\, C = \{ (\sigma_1,v_{\sigma_1}), \dots, (\sigma_{k},v_{\sigma_k}) \} \,\forall\, v_{\sigma_l} \in A \,\forall\, l \in [1,k]\}$ for $k \in [1,N]$ and $\mathcal{C}_{\sigma}(0) := \{ \{ \} \}$.\par
The value selection is pattern-based in analogy to \gls{CWFC}, which together with the identifier selection leads to \cref{eqn:pCQ}. Given $\sigma$ and $R \in \mathcal{R}^{\mathrm P}$, the task is to prepare $\ket{\Psi}$ from \cref{eqn:pCQ} constructively with quantum gates. Let $\ket{\Psi} := \operator{U}_N \cdots \operator{U}_1 \ket{0}$ as in \cref{fig:qwfccircuit}(a). Then,
\begin{equation} \label{eqn:Uk}
	\operator{U}_k := \left[ \operator{U}_k^{\mathrm{coin}} + \operator{U}_k^{0} \right] \otimes \bigotimes_{l=k+1}^N \mathbbm{1}_{\sigma_{l}} \in \mathcal{H}
\end{equation}
for $k \in [1,N]$, which contains $\mathbbm{1}_i \in \mathcal{H}_i$, the unit operator for $i \in [1,N]$ and two other operators.\par
First, the conditional initialization operator
\begin{equation} \label{eqn:Ukcoin}
	\operator{U}_k^{\mathrm{coin}} = \sum_{C' \in \mathcal{C}_{\sigma}(k)} \left[ \operator{U}_k^{\mathrm{ctrl}}(C') \otimes \operator{U}_k^{\mathrm{init}}(C') \right],
\end{equation}
where $\operator{U}_k^{\mathrm{ctrl}}(C') := \bigotimes_{(s,v) \in L_k(C')} \ket{v-1} \bra{v-1}_s \in \bigotimes_{l=1}^{k-1} \mathcal{H}_{\sigma_{l}}$ denotes the control operator, and $\operator{U}_k^{\mathrm{init}}(C') := \ket{C'} \bra{0}_{\sigma_k} \in \mathcal{H}_{\sigma_{k}}$ the initialization operator with $\ket{C'}_{\sigma_k} := \sum_{\nu \in A} \sqrt{p_k^{\mathrm{v,P}}(\nu | \sigma_k, C')} \ket{ \nu-1 }_{\sigma_k} \in \mathcal{H}_{\sigma_k}$. The summation in $\operator{U}_k^{\mathrm{ctrl}}(C')$ runs over all adjacent segments $L_k(C') := \{ (s,v) \,|\, (s,v) \in C' \,\land\, \exists\, P \in R_P\exists\, d \in P_d : \alpha^d_{\sigma_k s} = 1 \} \subseteq C'$, where $R_P := \{ P_1, \dots, P_m \}$ stands for the set of patterns in $R \in \mathcal{R}^{\mathrm P}$ with $P_i \in \mathcal{P}$ for $i \in [1,m]$ and $P_d := \{ d_1, \dots, d_n \}$ stands for the set of directions in the pattern $P \in \mathcal{P}$, \cref{eqn:P}, with $d_j \in [1,D]$ for $j \in [1,n]$. Choosing $C'$ instead of $L_k(C')$ in the summation has no effect on $p_k^{\mathrm{v,P}}(\nu | i, C')$ and thus $\ket{C'}_{\sigma_k}$.\par 
In iteration $k$, only the qubits in $\mathcal{Q}_{k-1}^+$ with $\mathcal{Q}_k^+ := \{ q \,|\, q = q_k^i \in \mathcal{Q}_{\sigma_k} \,\forall\, i \in [1,q] \,\land\, \sigma_k \in L_k^+ \}$ are affected by $\operator{U}^{\mathrm{ctrl}}(C')$ for some $C' \in \mathcal{C}_{\sigma}(k)$, where $L_k^+ := \{ s \,|\, \exists\, C' \in \mathcal{C}_{\sigma}(k) \,:\, s \in S^+(L_k(C')) \}$ denotes the affected segments for $k=[1,N]$. Here, $\mathcal{Q}_k^+ \subseteq \{\mathcal{Q}_1,\dots,\mathcal{Q}_k\}$ and $L_k^+ \subseteq \{\sigma_1,\dots,\sigma_k\}$, respectively.\par
The second operator in \cref{eqn:Uk} is the neutral operator
\begin{equation} \label{eqn:Uk0}
	\operator{U}_k^{0} = \sum_{C' \in \mathcal{C}_{\sigma}(k)} \left[ \operator{U}_k^{\mathrm{ctrl}}(C') \otimes \operator{U}_k^{\mathrm{init},0} + \operator{U}_k^{\mathrm{ctrl},0}(C') \right].
\end{equation}
Initially, all qubits $\mathcal{Q}_k$ are in the ground state. Therefore, $\operator{U}_k^{\mathrm{init},0} := \sum_{v \in [2,W]} \ket{v-1} \bra{v-1}_{\sigma_k} \in \mathcal{H}_{\sigma_{k}}$ vanishes. Furthermore, $\operator{U}_k^{\mathrm{ctrl},0}(C') := \left[ \bigotimes_{l=1}^{k-1} \mathbbm{1}_{\sigma_{l}} - \operator{U}_k^{\mathrm{ctrl}}(C') \right] \otimes \mathbbm{1}_{\sigma_{k}} \in \bigotimes_{l=1}^{k} \mathcal{H}_{\sigma_{l}}$ does not change the state by construction.\par
To realize $\operator{U}_k$ with quantum gates for a given $k \in [1,N]$, only $\operator{U}_k^{\mathrm{coin}}$ has to be considered (because all other parts of $\operator{U}_k$ have no effect on the resulting state), which consists of a sequence of joint pairs $\operator{U}_k^{\mathrm{ctrl}}(C') \otimes \operator{U}_k^{\mathrm{init}}(C')$, one for each $C' \in \mathcal{C}_{\sigma}(k)$. Each of these joint pairs represent a conditional loading of a probability distribution to realize $\ket{C'}_{\sigma_k}$, which is a well-known task that may, however, require an exponential number of gates~\cite{dasgupta2022}.\par 
In the original \gls{WFC}, a pattern conflict leads to a restart of the algorithm. This can also be realized in \gls{QWFC} by adding an additional ``conflict detection qubit'' that stores this information, making \ref{ps1}, \ref{ps2}, \ref{pv1}, and \ref{pv2} obsolete. This optional extension is not further discussed.

\subsection{Technical Details of HWFC}
Presume $H$ partitions such that partition $h \in [1,H]$ contains all of the segment identifiers from the set $S^h \subset [1,N]$ with $\bigcup_{h=1}^H S^h = [1,N]$ and $\bigcap_{h=1}^H S^h = \{\}$. The already sampled sub-contents $C^1,\dots,C^{h-1}$ are used as an additional constraint, which has to be taken account in the quantum circuit by replacing $\mathcal{C}_{\sigma}(k)$ in \cref{eqn:Ukcoin,eqn:Uk0} with $\mathcal{C}^h_{\sigma}(k) := \{ C \,|\, C = \bigcup_{i=1}^{h-1} C^i \cup C' \,\forall\, C' \in \mathcal{C}_{\sigma}(k) \}$.
	
	\section{ACKNOWLEDGMENTS}
Parts of this research have been funded by the Ministry of Science and Health of the State of Rhineland-Palatinate (Germany) as part of the project \emph{AnQuC}.	
	
	\def\refname{REFERENCES}
	
	\bibliographystyle{unsrt}
	{\normalsize\bibliography{refs}}

\end{document}